\documentclass[amsmath,amssymb, preprint]{revtex4-1}

\usepackage{dcolumn}
\usepackage{bm}

\usepackage[latin1]{inputenc}
\usepackage[english]{babel}

\usepackage{graphicx}
\usepackage{braket}
\usepackage[pdftex,plainpages=false,colorlinks,hyperindex,bookmarksopen,linkcolor=red,citecolor=blue,urlcolor=blue]{hyperref}

\usepackage{mathrsfs}

\usepackage{epstopdf}

\usepackage{color}

\newcommand{\be}{\begin{eqnarray}}
\newcommand{\ee}{\end{eqnarray}}

\newcommand{\R}{\mathbb{R}}  
\newcommand{\C}{\mathbb{C}} 
\newcommand{\N}{\mathbb{N}} 
\def\eg{{\it e.g. }} 
\def\ie{{\it i.e. }}
\newcommand{\D}{\mathcal{D}}

\def\eg{{\it e.g.}\ }
\def\ie{{\it i.e.}\ }

\begin{document}

\title{Dispersion relations for the time-fractional\\ Cattaneo-Maxwell heat equation}

 \author{Andrea Giusti}
		\affiliation{Department of Physics $\&$ Astronomy,\\
		University of Bologna and INFN. Via Irnerio 46, Bologna, ITALY\\
		 and\\ 
	    	 Arnold Sommerfeld Center,\\
	    	 Ludwig-Maximilians-Universit\"at, 
	    	 Theresienstra{\ss}e~37, 80333 M\"unchen, GERMANY.}	
 		\email{agiusti@bo.infn.it}
 

    \date  {\today}

\begin{abstract}
In this paper, after a brief review of the general theory of dispersive waves in dissipative media, we present a complete discussion of the dispersion relations for both the ordinary and the time-fractional Cattaneo-Maxwell heat equations. Consequently, we provide a complete characterization of the group and phase velocities for these two cases, together with some non-trivial remarks on the nature of wave dispersion in fractional models.

\vskip 2.0 cm

\begin{center}
\textit{Journal of Mathematical Physics} \textbf{59}, 013506 (2018)\\
DOI: \href{https://doi.org/10.1063/1.5001555}{10.1063/1.5001555}
\end{center}
\end{abstract}


\maketitle

\section{Introduction} \label{Sec-Intro}
	A precise mathematical description of the propagation of heat waves lays at the very foundations of electro-mechanical systems involved in modern technology. Historically, the first proposal for a mathematical model for thermal conduction was introduced by J.~Fourier by means of an \textit{ad hoc} argument that allowed to properly explain some experimental results. Specifically, Fourier's law states that the local heat flux density $\textbf{q}$ is equal to the product of thermal conductivity $\kappa$ and the opposite of the temperature gradient, $- \nabla T$, \ie
\be 
\textbf{q} = - \kappa \, \nabla T \, .
\ee
	Then, in the late 40s, C.~Eckart showed that such a law can actually be recovered from the theory of non-equilibrium thermodynamics, see \cite{E-1, E-2, Ruggeri}.

Following the discussion in \cite{Ruggeri}, if we consider a given fluid, then its thermodynamic behaviour is completely characterized by five classical fields, \ie mass density $\rho$, velocity $\textbf{v}$ and temperature $T$. Now, let us consider a fluid with constant mass density and specific heat $\varepsilon$, then the amount of heat in a region $\Omega \subset \R^3$, with a boundary $\partial \Omega$, is given by
\be 
Q_\Omega = \int _\Omega \varepsilon \, \rho \, T \, \mathrm{d} ^3 x  \, .
\ee
If $T = T(t, \textbf{x})$ is sufficiently smooth, then
\be \label{eq-derQ}
\frac{\partial Q_\Omega}{\partial t} = \int _\Omega \varepsilon \, \rho \, \frac{\partial T}{\partial t} \, \mathrm{d} ^3 x \, .
\ee
If no work is done and there are neither heat sources nor sinks, Fourier's law then allows us to compute the change in heat energy in the region $\Omega$. Indeed, it is accounted for its entirety by the flux of heat across the boundaries, namely
$$ \frac{\partial Q_\Omega}{\partial t} = \int _{\partial \Omega} \kappa \, \nabla T \cdot \textbf{n} \, \mathrm{d}^2 x \, .$$
Applying the divergence theorem one then finds
\be \label{eq-Gauss}
\frac{\partial Q_\Omega}{\partial t} = \int _\Omega \texttt{div} \, (\kappa \, \nabla T) \, \mathrm{d}^3 x = 
\int _\Omega \kappa \, \Delta T \, \mathrm{d}^3 x \, ,
\ee
where $\Delta \equiv \texttt{div} \, \nabla$ is the Laplacian operator.
Finally, equating Eq.~\eqref{eq-derQ} and Eq.~\eqref{eq-Gauss} we get the \textbf{parabolic Heat equation}:
\be \label{eq-diff}
\frac{\partial T}{\partial t} = \frac{\kappa}{\varepsilon \, \rho} \, \Delta T \, .
\ee
which is also known as \textbf{diffusion equation}.

	However, despite the extraordinary accordance between Fourier's law and most of the experimental data, Eq.~\eqref{eq-diff} leads to a major conceptual paradox. Indeed, the solution of an initial value problem on $\R^3$ for Eq.~\eqref{eq-diff} is given by
	\be 
	T(t, \textbf{x}) = 
	\frac{1}{(4 \, \pi \, D \, t)^{3/2}} \int _{\R ^3} T (0, \textbf{y}) \, \exp \left( - \frac{|\textbf{x} - \textbf{y}|^2}{4 \, D \, t} \right) \, \mathrm{d} ^3 y \, ,
	\qquad
	D \equiv \frac{\kappa}{\rho \, \varepsilon} \, ,
	\ee
which clearly tells us that in a model for heat waves based on the diffusion equation the temperature spreads throughout the whole space infinitely fast. Therefore, this formulation for the theory of heat propagation violates the causality principle by predicting an infinite propagation speed.

	A reformulation of thermodynamics involving a finite speed propagation for heat waves, \ie in terms of hyperbolic partial differential equations, was therefore needed \cite{Ruggeri}. This line of research was started in the 30s by B.~I.~Davydov\cite{Bakunin} and then independenly treated by P.~Vernotte\cite{Vernotte} and C.~Cattaneo\cite{Cattaneo}, in the 50s, whose work led to some major steps forward in the resolution of this paradoxical nature of the classical theory for heat conduction.

	Cattaneo's idea was fairly simple: in order to restore causality one has to modify the constitutive equation for heat conduction, \ie Fourier's law. The Cattaneo-Maxwell law \cite{CM-law, Pipkin, Preziosi, Straughan} is the most known among the various modifications of Fourier's law and takes the form 
	\be \label{Cattaneo}
	\textbf{q} + \tau \, \frac{\partial \textbf{q}}{\partial t} = - \kappa \, \nabla T \, , 
	\ee
	where $\tau$ is the so called \textit{relaxation time}, and it leads to a modified version of the heat equation,
	\be \label{CM}
	\tau \, \frac{\partial ^2 T}{\partial t^2} + \frac{\partial T}{\partial t} = D \, \Delta T \, , 
	\ee
	which is known as the \textbf{Cattaneo-Maxwell heat conduction equation} (or, alternatively, as Cattaneo's heat conduction law or Cattaneo heat equation), which represents a particular realization of the telegraph equation (see \eg Ref.\cite{Mainardi-book}), where $[\tau]=\mbox{time}$, $[D]=\mbox{length} ^2 \, \cdot \mbox{time}^{-1}$.

	In the last few years, the quest for potential applications of fractional calculus \cite{Mainardi-1997, Mainardi-book, Kilbas, Giusti-Comment} in biology \cite{IC-AG-FM-ZAMP, AG-FM_MECC16, Silvia-1, Silvia-2}, thermodynamics \cite{Fabrizio, GaGiMa, Garra, Garcia}, viscoelasticity \cite{IC-AG-FM-Bessel, AG-FCAA-2017, AG-FM-EPJP, Masina} has been attracting much attention in the mathematical community. Nevertheless, despite this growing interest, not much work has been done in the study of dispersion relations for fractional models for wave propagation. The aim of this paper is, therefore, to provide a few remarks on fractional dispersion relations by analysing the specific example of the causal heat diffusion.

	The paper is organized as follows:
	
	In Section~\ref{sec-2} we present a brief discussion on general aspects of linear dispersive waves with dissipation.
	
	In Section~\ref{sec-3}, we review some of the results presented in Ref. \cite{Vitokhin} by setting up the problem following the general formalism discussed in Section~\ref{sec-2}. Specifically, we discuss in full details the dispersion law for the ordinary Cattaneo diffusion law and present some further remarks concerning the nature of the (anomalous) dispersion due to the structure of this wave equation.
	
	Then, in Section~\ref{sec-4}, we perform a complete study of the dispersion relation for the fractional Cattaneo-Maxwell heat conduction equation.
	
	Finally, we conclude the paper by comparing the results obtained in Section~\ref{sec-3} and Section~\ref{sec-4} and providing some hints for future research.

\section{Linear dispersive waves with dissipation} \label{sec-2}
	Linear dispersive waves arise in systems whose dynamics is governed by a set of linear equations, subject to linear initial and boundary conditions. The attribute dispersive further implies the existence of a non-trivial relationship between the wave number $k$ and the angular frequency $\omega$ in the elementary sinusoidal solution, known as the \textit{dispersion law}.

	Let $\varphi (t, x)$ be the wave function for a $(1+1)$-dimensional system, then the elementary sinusoidal ansatz reads
	\be 
	\varphi (t, x) = \texttt{Re} \left\{ \Phi \, e^{i (k \, x - \omega \, t)} \right\}
	\ee
	where $\Phi$ is known as the \textit{complex amplitude} and the parameters $k$ and $\omega$ satisfy the dispersion relation
	\be \label{dispersion}
	\mathcal{D} (\omega, k) = 0 \, ,
	\ee
	where $\mathcal{D}$ is a suitable real function of $\omega$ and $k$. Such an equation is, in general, solved by certain $\omega, k \in \C$.

	Let us assume that (\ref{dispersion}) can be solved explicitly in terms of a real parameter ($k$ or $\omega$) by means of complex valued branches:
	\be
	\overline{\omega} _\ell (k) & \in & \C \, , \,\,\,\, k \in \R \, ,\\
	\overline{k} _m (\omega) & \in & \C \, , \,\,\,\, \omega \in \R \, ,
	\ee	
	where $\ell , \, m$ are two positive integers called mode indices. These branches are then related to the \textit{normal mode solutions} of the dynamical equations for the physical system, \ie
	\be 
	\varphi _\ell (t, x; k) &=& \hbox{Re} \left\{ \Phi _\ell (k) \, \exp\left[ i ( k \, x - \overline{\omega} _\ell \, t) \right] \right\} \, , \label{conf-1} \\ 
	\varphi _m (t, x; \omega) &=& \hbox{Re} \left\{ \Phi _\ell (\omega) \exp\left[ i ( \overline{k} _m \, x - \omega \, t ) \right] \right\} \, . \label{conf-2}
	\ee
	
	Hence, the two types of normal modes expansions read
\be \label{eq-exp-1}
\varphi (t, x) = \sum _\ell \int _{\mathcal{C} _\ell} \alpha _\ell (k) \, \varphi _\ell (t, x; k) \, dk \, ,
\ee	
\be \label{eq-exp-2}
\varphi (t, x) = \sum _m \int _{\mathcal{C} _m} \beta _m (\omega) \, \varphi _m (t, x; \omega) \, d\omega \, , 
\ee	
where $\mathcal{C}$ is either the real line or, in case of singularities on it, a parallel line properly chosen to ensure the convergence of the integral. Besides, $\alpha _\ell (k)$ and $\beta _m (\omega)$ are complex valued functions to be determined in accordance with the initial or boundary conditions.	

	From now on we will omit the mode labels, for sake of clarity. 
	
	Let us now define, for the two cases (\ref{conf-1}) and (\ref{conf-2}) respectively, the phase velocity as
	\be 
	v_p (k) := \frac{\texttt{Re}\, \overline{\omega} (k)}{k} \, , \label{eq-vp-gen} \\
	v_p (\omega) := \frac{\omega}{\texttt{Re}\, \overline{k} (\omega)} \, .
	\ee
	Furthermore, let us also define, for both cases, the group velocity as
	\be 
	v_g (k) :=  
	\frac{\partial}{\partial k} \, \texttt{Re} \,\overline{\omega} (k) \, .
	\ee

	Now, from Eq.~\eqref{conf-1} it is easy to see that $\varphi _\ell (t, x; k)$ is sinusoidal in space with a \textit{wavelength} $\lambda = 2 \, \pi / k$. However, the sinusoidal nature in time is not guaranteed given that
	\be 
	\overline{\omega} = \omega _r + i \, \omega _i
	\ee
	where $\omega _r = \texttt{Re} \, \overline{\omega}$ and $\omega _i = \texttt{Im} \, \overline{\omega}$ are two real functions of $k$, then Eq.~\eqref{conf-1} can be rewritten as
	\be 
	\varphi _\ell (t, x; k) &=& e^{- \gamma (k) \, t} \, \texttt{Re} \left\{ \Phi _\ell (k) \, \exp\left[ i \, k \, ( x - v_p (k) \, t ) \right] \right\} \, ,
	\ee
	where, if $\omega _i \leq 0$, $\gamma (k) = - \omega _i (k)$ is known as the \textit{time-damping factor}.

	Analogously, from Eq.~\eqref{conf-2} it is easy to see that $\varphi _m (t, x; \omega)$ is sinusoidal in time with a \textit{period} $T = 2 \, \pi / \omega$. Similarly, the sinusoidal nature in space is not guaranteed given that
	\be 
	\overline{k} = k _r + i \, k _i
	\ee
	where $k _r = \texttt{Re} \, \overline{k}$ and $k _i = \texttt{Im} \, \overline{k}$ are two real functions of $\omega$, then Eq.~\eqref{conf-2} can be rewritten as
	\be 
	\varphi _m (t, x; \omega) &=& e^{- \delta (\omega) \, t} \, 
	\texttt{Re} \left\{ \Phi _m (\omega) \, \exp\left[ i \, \omega \, \left( \frac{x}{v_p (\omega)} - t \right) \right] \right\} \, ,
	\ee
	where, if $k _i \geq 0$, $\delta (\omega) = k _i (\omega)$ is known as the \textit{space-damping factor}.

	For sake of brevity, in the following we shall study only the complex frequency branch, \ie $\overline{\omega} _\ell (k)$, for both the Cattaneo-Maxwell heat conduction equation and its time fractional counterpart.

\section{Dispersion relation for the Cattaneo-Maxwell heat equation} \label{sec-3}
	Let us start off with a study of the dispersion relation for the hyperbolic heat conduction equation \eqref{CM}, in analogy with the analysis in Ref. \cite{Vitokhin}.
	
	Hence, let us focus ourself on the $(1+1)$-dimensional Cattaneo-Maxwell heat conduction equation, that reads
	\be \label{CM-1d}
	\tau \, \frac{\partial ^2 T}{\partial t^2} + \frac{\partial T}{\partial t} = D \, \frac{\partial ^2 T}{\partial x^2} \, , 
	\ee
	with $[\tau]=\mbox{time}$, $[D]=\mbox{length} ^2 \, \cdot \mbox{time}^{-1}$.	
	
	Following the discussion presented in Section~\ref{sec-2} for the complex frequency decomposition, if we plug into Eq.~\eqref{CM-1d} an ansatz of the form
	\be 
	T(t, x) \sim \exp\left[i (k \, x  - \overline{\omega} \, t)\right] \, ,
	\ee
	with $\overline{\omega} = \omega_r + i \, \omega_i$, then we get the corresponding dispersion relation (in the complex frequency branch)
	$$ \tau \, \overline{\omega} ^2 + i \, \overline{\omega} - D \, k^2 = 0 \, . $$
	
	Solving the latter with respect to $\overline{\omega}$ we get
	\be 
	\overline{\omega} (k) = \frac{-i \pm \sqrt{4 \, \tau \, D \, k^2 -1 }}{2 \, \tau} \, ,
	\ee
	from which we can infer that (choosing the solution with the positive sign in front of the square root, \ie the positive branch)
	\be 
	\omega _r (k) =	
	\left\{
	\begin{aligned}
	& 0 \, , \quad 0 \leq k \leq 1/\sqrt{4 \, \tau \, D} \, ,\\
	& \sqrt{\frac{D \, k^2}{\tau} - \frac{1}{4 \, \tau ^2}} \, , \quad k > 1/\sqrt{4 \, \tau \, D} \, .
	\end{aligned}
	\right.
	\ee
	
	\be 
	\omega _i (k) =	
	\left\{
	\begin{aligned}
	& \frac{-1 + \sqrt{1 - 4 \, \tau \, D \, k^2}}{2 \, \tau} \, , \quad 0 \leq k \leq 1/\sqrt{4 \, \tau \, D} \, ,\\
	& - \frac{1}{2 \, \tau} \, , \quad k > 1/\sqrt{4 \, \tau \, D} \, .
	\end{aligned}
	\right.
	\ee
	
	If we consider the case in which $D, \tau \gg 1$ then Eq.~\eqref{CM} can be rewritten as a d'Alembert equation with typical speed $c = \sqrt{D/\tau}$.	
	
	It is also worth remaking that, in this case, the time-damping coefficient $\gamma (k) = - \omega _i (k)$ grows with $k$ in $0 \leq k \leq 1/\sqrt{4 \, \tau \, D}$ and then it stabilizes to the constant value $\gamma (k) = 1 /2 \, \tau$ for $k > 1/\sqrt{4 \, \tau \, D}$.
	
	\begin{figure}[h!] \label{fig-1}
	\centering
	\includegraphics[scale=0.32]{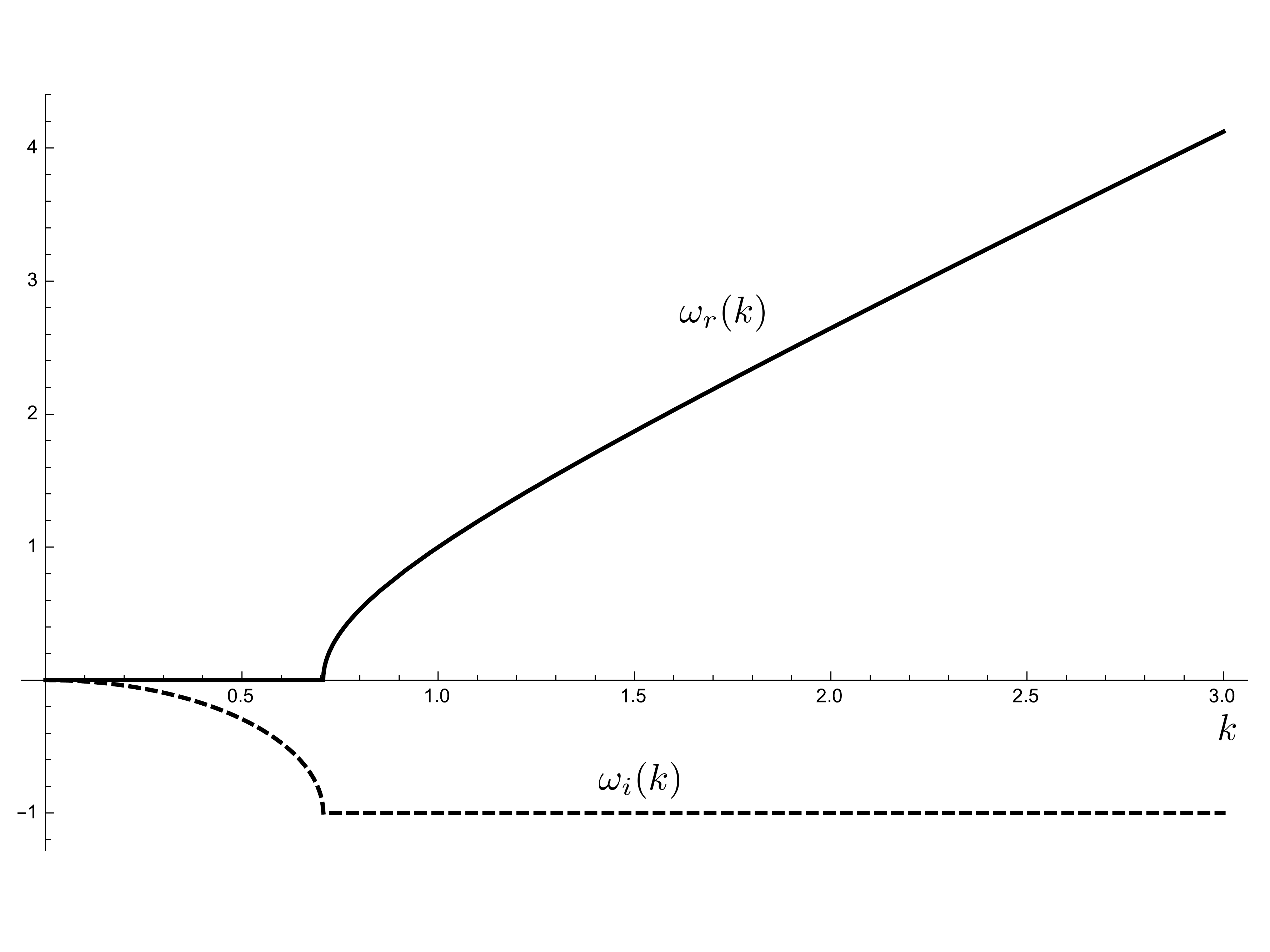}
	\caption{Complex frequency as a function of the wavenumber for the ordinary Cattaneo-Maxwell heat equation. ($\tau = 0.5$ and $D = 1$)}
	\end{figure}	
	
\newpage	
	
The phase and group velocities are then easily computed, \ie
\be
v_p (k) = \frac{\omega_r (k)}{k} =  
	\left\{
	\begin{aligned}
	& 0 \, , \quad 0 \leq k \leq 1/\sqrt{4 \, \tau \, D} \, ,\\
	& \sqrt{\frac{D}{\tau} - \frac{1}{4 \, \tau ^2 \, k^2}} \, , \quad k > 1/\sqrt{4 \, \tau \, D} \, .
	\end{aligned}
	\right.
\ee

\be
v_g (k) = \frac{\partial \omega_r (k)}{\partial k} =  
	\left\{
	\begin{aligned}
	& 0 \, , \quad 0 \leq k \leq 1/\sqrt{4 \, \tau \, D} \, ,\\
	& \frac{2 \, D \, k}{\sqrt{4 \, D \, \tau \, k^2 - 1}} \, , \quad k > 1/\sqrt{4 \, \tau \, D} \, .
	\end{aligned}
	\right.
\ee	

	\begin{figure}[h!] \label{fig-2}
	\centering
	\includegraphics[scale=0.32]{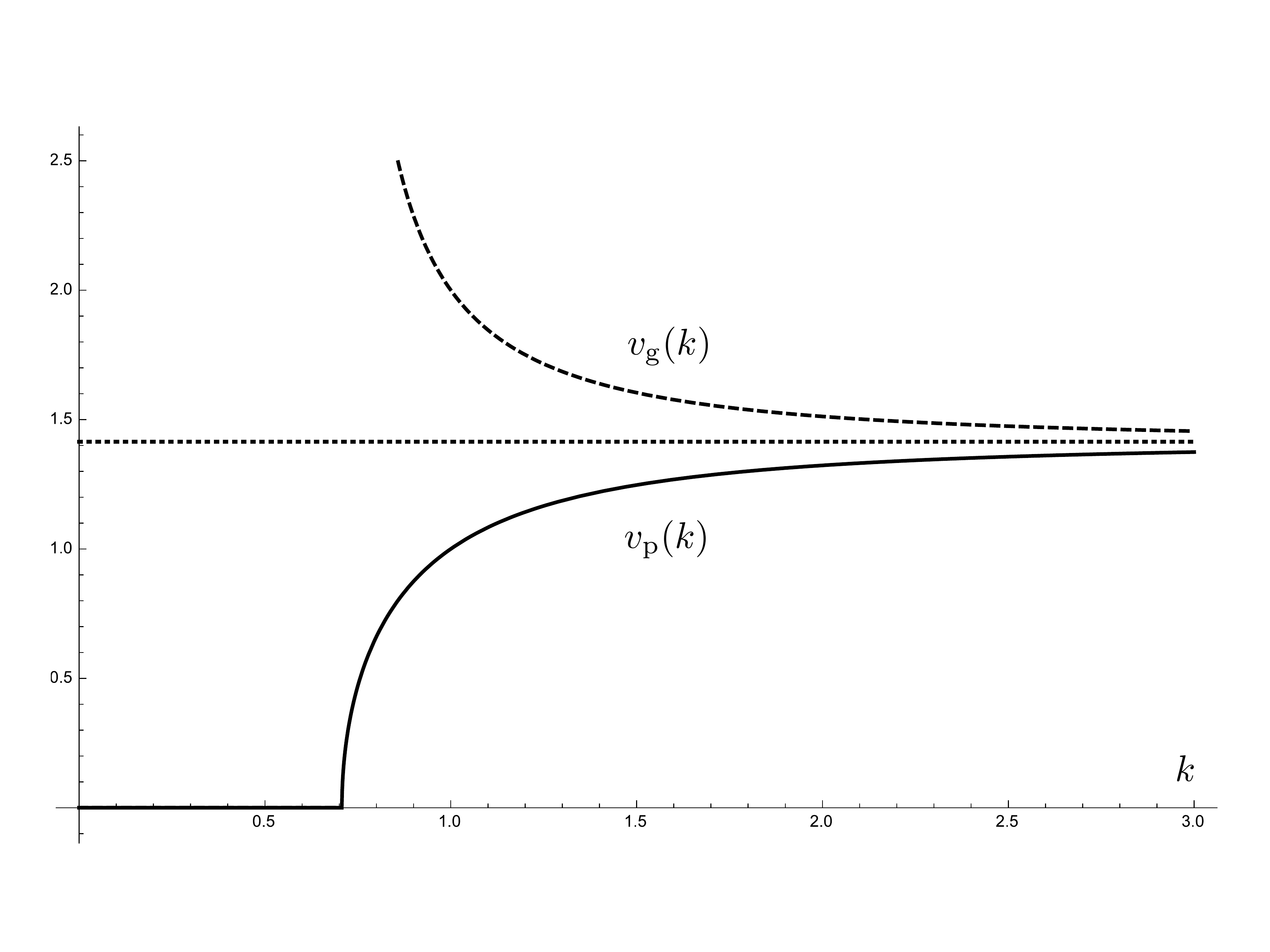}
	\caption{Phase and group velocities as functions of the wavenumber for the ordinary Cattaneo-Maxwell heat equation. ($\tau = 0.5$ and $D = 1$)}
	\end{figure}

	From these relations and plots one can easily infer that:
	\begin{itemize}
	\item In a regime in which $k \gg 1$ the system tends to recover a sort of dispersion-less behaviour, with characteristic velocity $c = \sqrt{D/\tau}$,  even thought a certain degree of dispersion is still present;
	\item Nonetheless, $\forall k > 1/\sqrt{4 \, \tau \, D}$ the system features an \textbf{anomalous dispersion}, \ie $v_g (k) > v_p (k)$.
	
	Indeed,
	\be \notag
	v_g (k) - v_p (k) &=& \frac{2 \, D \, k}{\sqrt{4 \, D \, \tau \, k^2 - 1}} - \sqrt{\frac{D}{\tau} - \frac{1}{4 \, \tau ^2 \, k^2}} =\\ \notag
	&=& \frac{c}{\sqrt{1 - \frac{1}{4 \, c^2 \, \tau^2 \, k^2}}} - c \, \sqrt{1 - \frac{1}{4 \, c^2 \, \tau^2 \, k^2}} =\\ \notag
	&=& \frac{1}{2 \, \tau \, k \, \sqrt{4 \, c^2 \, \tau^2 \, k^2 - 1}} > 0  \, .
	\ee
	\end{itemize}

\section{Dispersion relation for the time-fractional Cattaneo-Maxwell heat equation} \label{sec-4}
	Let us now turn our attention to the fractional version of Eq.~\eqref{CM-1d}. Some general aspects of the fractionalization of Fourier's law and of the corresponding fractional heat conduction equation have already been studied in the literature, see \eg \cite{Fabrizio, Garra, Qi}. In particular, it is worth recalling the seminal paper\cite{Metzler} by Compte and Metzler on the application of some fractional generalizations of the Cattaneo equation to describe anomalous transport processes.
	
	Following the discussion in Ref. \cite{Qi}, it is important to stress that a fractionalization in time corresponds to the introduction of some extra memory effects in the system's dynamics, whereas a fractionalization in space induces non-localities in the system. In this paper we are particularly interested in the study of the emergence of memory effects in the process of causal heat conduction, therefore we will neglect the contribution of non-local terms. 
	
	Hence, the time-fractional Cattaneo-Maxwell heat equation, see Ref. \cite{Qi}, with $0 < \alpha < 1$ is given by
	\be \label{FCM}
	\tau^\alpha \, \textbf{D} ^{2\alpha} \, T + \textbf{D} ^{\alpha} \, T = \overline{D} \, \frac{\partial ^2 T}{\partial x^2} \, , 
	\ee
	with $[\tau]=\mbox{time}$, $[\overline{D}]=\mbox{length} ^2 \, \cdot \mbox{time}^{-\alpha}$ and where
	$$ \textbf{D} ^{\alpha} f (t) \equiv {}_{-\infty} {}^C \D_t ^\alpha f(t) := \frac{1}{\Gamma (1 -\alpha)} 
	\int _{- \infty} ^t \frac{f' (\tau)}{(t - \tau) ^{\alpha}} \, d \tau \, ,$$
	is the time-fractional Caputo derivative.
	
	Notice that with this definition of fractional derivative (with $t_{initial} = -\infty$) we are implicitly assuming that we will be dealing with the dispersion of waves which are the result of perturbations that occurred way before the time of observation. 
	
	Now, recalling the Fourier transform for $\textbf{D} ^{\alpha} f (t)$ (see \eg Ref. \cite{Kilbas}), the dispersion relation corresponding to the wave equation \eqref{FCM} reads (complex frequency branch)
	\be 
	\tau ^\alpha \, (- i \, \overline{\omega}) ^{2 \alpha} + (- i \, \overline{\omega}) ^{\alpha} + \overline{D} \, k^2 = 0 \, ,
	\ee
whose solutions are given by
	\be 
	(- i \, \overline{\omega}) ^{\alpha} =
	\frac{-1 \pm \sqrt{1 - 4 \, \tau ^\alpha \, \overline{D} \, k^2}}{2 \, \tau ^\alpha} \, ,
	\ee
	thus,
	\be \label{Complex-Omega}
	\overline{\omega} ^\alpha = i^\alpha \, \frac{-1 \pm \sqrt{1 - 4 \, \tau ^\alpha \, \overline{D} \, k^2}}{2 \, \tau ^\alpha} \, .
	\ee

	In order to explicitly compute $\overline{\omega} (k)$ it is convenient to study Eq.~\eqref{Complex-Omega} by means of the exponential representation of complex numbers, \ie plugging
	$$ \overline{\omega} = \left| \overline{\omega} \right| \, e^{i \, \theta} \, , \qquad \left| \overline{\omega} \right| > 0, \,\, -\pi < \theta < \pi $$
into Eq.~\eqref{Complex-Omega}.

	Hence, Eq.~\eqref{Complex-Omega} now reads
	\be \label{Eq-Complex-Omega-2}
	\left| \overline{\omega} \right| ^\alpha \, e^{i \, \alpha \theta} = 
	e^{i \, \alpha \pi /2} \, \frac{-1 \pm \sqrt{1 - 4 \, \tau ^\alpha \, \overline{D} \, k^2}}{2 \, \tau ^\alpha} \, .
	\ee

	Then, to solve the latter we have to distinguish two cases, \ie
	\begin{itemize}
	\item[(a)] $1 - 4 \, \tau ^\alpha \, \overline{D} \, k^2 < 0$, which gives us an extra contribution to the imaginary part of $\overline{\omega}$;
	\item[(b)] $1 - 4 \, \tau ^\alpha \, \overline{D} \, k^2 \geq 0$.
	\end{itemize}
	
	\vskip 0.3 cm

\noindent \textbf{Case (a)}. If $1 - 4 \, \tau ^\alpha \, \overline{D} \, k^2 < 0$ then, setting $A \equiv \sqrt{\left|1 - 4 \, \tau ^\alpha \, \overline{D} \, k^2\right|}$, Eq.~\eqref{Eq-Complex-Omega-2} reads
\be 
\left| \overline{\omega} \right| ^\alpha \, e^{i \, \alpha \theta} = 
	e^{i \, \alpha \pi /2} \, \frac{-1 \pm i \, A}{2 \, \tau ^\alpha} \, .
\ee
 
	Rewriting the second term in the l.h.s. in the exponential form, \ie
	$$ \frac{-1 \pm i \, A}{2 \, \tau ^\alpha} = \rho \, e^{i \, \psi} \, , $$
	where
	$$ \rho = \left| \frac{-1 \pm i \, A}{2 \, \tau ^\alpha} \right| = \sqrt{\frac{\overline{D} \, k^2}{\tau ^\alpha}} \, , \quad \psi = \mp \arctan A \,  $$

	Therefore,
	$$ |\overline{\omega}| ^\alpha = \sqrt{\frac{\overline{D} \, k^2}{\tau ^\alpha}} \, , \qquad \alpha \, \theta = \frac{\alpha \, \pi}{2} + \psi \, , $$
from which we can infer that
\be \label{Case-a}
\boxed{
|\overline{\omega}| = \left( \frac{\overline{D} \, k^2}{\tau ^\alpha} \right)^{1/2 \alpha} \, , \qquad \theta = \frac{\pi}{2} \mp \frac{1}{\alpha} \,
\arctan \left( \sqrt{4 \, \tau ^\alpha \, \overline{D} \, k^2 - 1} \right)
} \, .
\ee

	Now, if we take profit of the trigonometric representation for complex numbers, one can easily deduce that
	\be 
	\omega _r (k) &=& |\overline{\omega}| \, \cos \theta =\\ \notag
	&=& \pm \left( \frac{\overline{D} \, k^2}{\tau ^\alpha} \right)^{1/2 \alpha}
	\, \sin \left[ \frac{1}{\alpha} \,\arctan \left( \sqrt{4 \, \tau ^\alpha \, \overline{D} \, k^2 - 1} \right) \right] \, ,
	\ee
where we have used $\cos \left(\frac{\pi}{2} \pm \varphi \right) = \mp \sin \varphi$.

	Analogously,
	\be 
	\omega _i (k) &=& |\overline{\omega}| \, \sin \theta =\\ \notag
	&=& \left( \frac{\overline{D} \, k^2}{\tau ^\alpha} \right)^{1/2 \alpha}
	\, \cos \left[ \frac{1}{\alpha} \,\arctan \left( \sqrt{4 \, \tau ^\alpha \, \overline{D} \, k^2 - 1} \right) \right] \, ,
	\ee
where we have taken advantage of the relation $\sin \left(\frac{\pi}{2} \pm \varphi \right) = \cos \varphi$.

\vskip 0.3 cm

\noindent \textbf{Case (b)}. Conversely, if $1 - 4 \, \tau ^\alpha \, \overline{D} \, k^2 \geq 0$ then Eq.~\eqref{Eq-Complex-Omega-2} reads
\be 
\left| \overline{\omega} \right| ^\alpha \, e^{i \, \alpha \theta} = 
	e^{i \, \alpha \pi /2} \, \frac{-1 \pm A}{2 \, \tau ^\alpha} \, .
\ee

Now, one can notice that
$$ -1 \pm A (k) < 0 \, , \quad \forall k \in \R \, , $$
that implies
$$ \rho = \left| \frac{-1 \pm A}{2 \, \tau ^\alpha} \right| \, , \qquad \psi = \frac{\alpha \, \pi}{2} + \pi \, ,  $$
from which one gets that
\be \label{Case-b}
\boxed{
|\overline{\omega}| = \frac{1}{\tau} \, \left| \frac{-1 \pm \sqrt{1 - 4 \, \tau ^\alpha \, \overline{D} \, k^2}}{2} \right| ^{1/\alpha} \, , \quad 
\theta = \frac{\pi}{2} + \frac{\pi}{\alpha}
} \, .
\ee
or, in a different form: $\overline{\omega} = |\overline{\omega}| \, i \, (-1)^{1/\alpha}$, which represents a purely imaginary if $\alpha = 1/2n$, with $n \in \N$.

	Furthermore, as in \textbf{Case (a)}, one can compute the expressions for the real and the imaginary parts of $\overline{\omega}$. Indeed, here we have that
	\be 
	\omega _r (k) = |\overline{\omega}| \, \cos \theta = 
	- \frac{1}{\tau} \, \left| \frac{-1 \pm \sqrt{1 - 4 \, \tau ^\alpha \, \overline{D} \, k^2}}{2} \right| ^{1/\alpha} \, 
	\sin \left( \frac{\pi}{\alpha} \right) \, ,
	\ee 
	\be 
	\omega _i (k) = |\overline{\omega}| \, \cos \theta =
	\frac{1}{\tau} \, \left| \frac{-1 \pm \sqrt{1 - 4 \, \tau ^\alpha \, \overline{D} \, k^2}}{2} \right| ^{1/\alpha} \, 
	\cos \left( \frac{\pi}{\alpha} \right) \, ,
	\ee 
where we have taken profit of the identities: $\cos \left(\frac{\pi}{2} \pm \varphi \right) = \mp \sin \varphi$ and  $\sin \left(\frac{\pi}{2} \pm \varphi \right) = \cos \varphi$. 

	\vskip 0.5 cm
	
	To sum up, we have
	\begin{equation*}
	\omega _r (k) =
	\left\{	
	\begin{aligned}
	& - \frac{1}{\tau} \, \left( \frac{1 \mp \sqrt{1 - 4 \, \tau ^\alpha \, \overline{D} \, k^2}}{2} \right) ^{1/\alpha} \, 
	\sin \left( \frac{\pi}{\alpha} \right) \, , \quad 0 \leq k \leq 1 / \sqrt{4 \, \overline{D} \, \tau ^\alpha} \, , \\
	& \pm \left( \frac{\overline{D} \, k^2}{\tau ^\alpha} \right)^{1/2 \alpha}
	\, \sin \left[ \frac{1}{\alpha} \,\arctan \left( \sqrt{4 \, \tau ^\alpha \, \overline{D} \, k^2 - 1} \right) \right] , \quad
	k > 1 / \sqrt{4 \, \overline{D} \, \tau ^\alpha} \, ,
	\end{aligned}
	\right.	
	\end{equation*}

	\begin{equation*}
	\omega _i (k) =
	\left\{	
	\begin{aligned}
	& \frac{1}{\tau} \, \left( \frac{1 \mp \sqrt{1 - 4 \, \tau ^\alpha \, \overline{D} \, k^2}}{2} \right) ^{1/\alpha} \, 
	\cos \left( \frac{\pi}{\alpha} \right) \, , \quad 0 \leq k \leq 1 / \sqrt{4 \, \overline{D} \, \tau ^\alpha} \, , \\
	& \left( \frac{\overline{D} \, k^2}{\tau ^\alpha} \right)^{1/2 \alpha}
	\, \cos \left[ \frac{1}{\alpha} \,\arctan \left( \sqrt{4 \, \tau ^\alpha \, \overline{D} \, k^2 - 1} \right) \right] , \quad
	k > 1 / \sqrt{4 \, \overline{D} \, \tau ^\alpha} \, ,
	\end{aligned}
	\right.	
	\end{equation*}

	If we define $a = 1 / \sqrt{4 \, \overline{D} \, \tau ^\alpha}$ it is easy to see that
	\be \notag 
	\lim_{k \to a ^-} \omega _r (k) = - \frac{1}{2^\alpha \, \tau} \, \sin \left( \frac{\pi}{\alpha} \right) \quad \mbox{and} \quad 
	\lim_{k \to a ^+} \omega _r (k) = 0 \, .
	\ee
	Moreover,
	\be \notag 
	\lim_{k \to a ^-} \omega _i (k) = \frac{1}{2^\alpha \, \tau} \, \cos \left( \frac{\pi}{\alpha} \right) \quad \mbox{and} \quad 
	\lim_{k \to a ^+} \omega _i (k) = \frac{1}{2^\alpha \, \tau} \, .
	\ee
	Hence, $\omega _r (k)$ is continuous at $a$ if and only if $\alpha = 1/n$ with $n \in \N$ and $n > 1$. Besides, $\omega _i (k)$ is continuous at $a$ if and only if $\alpha = 1 / 2n$ with $n \in \N$ and $n \geq 1$.

	Therefore, this specific fractionalization of the Cattaneo-Maxwell diffusion law leads to a \textbf{jump discontinuity} in either the real or imaginary part of $\overline{\omega} (k)$ if $\alpha \neq 1/n$, with $n \in \N$ and $n > 1$. Furthermore, a continuous $\overline{\omega} (k)$, in both the real and imaginary parts, occurs only if $\alpha = 1/2n$, with $n \in \N$ and $n \geq 1$.

\begin{figure}[h!] 
\centering
\includegraphics[scale=0.3]{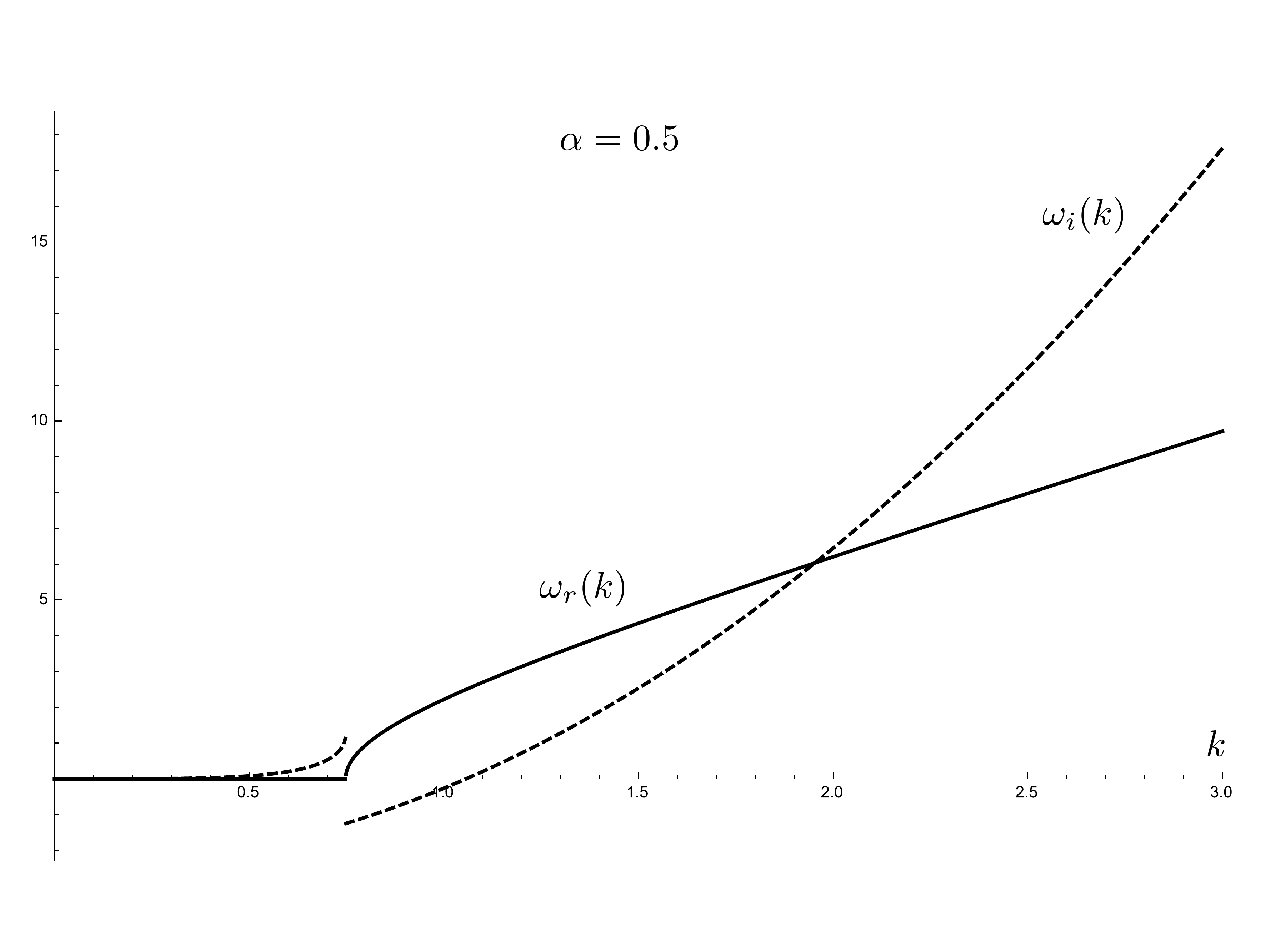}
\caption{Complex frequency as a function of the wavenumber for the fractional Cattaneo-Maxwell heat equation of order $\alpha = 1/2$. ($\tau = 0.2$ and $D = 1$) \label{cazzafa}}
\end{figure}

\begin{figure}[h!]
\centering
\includegraphics[scale=0.3]{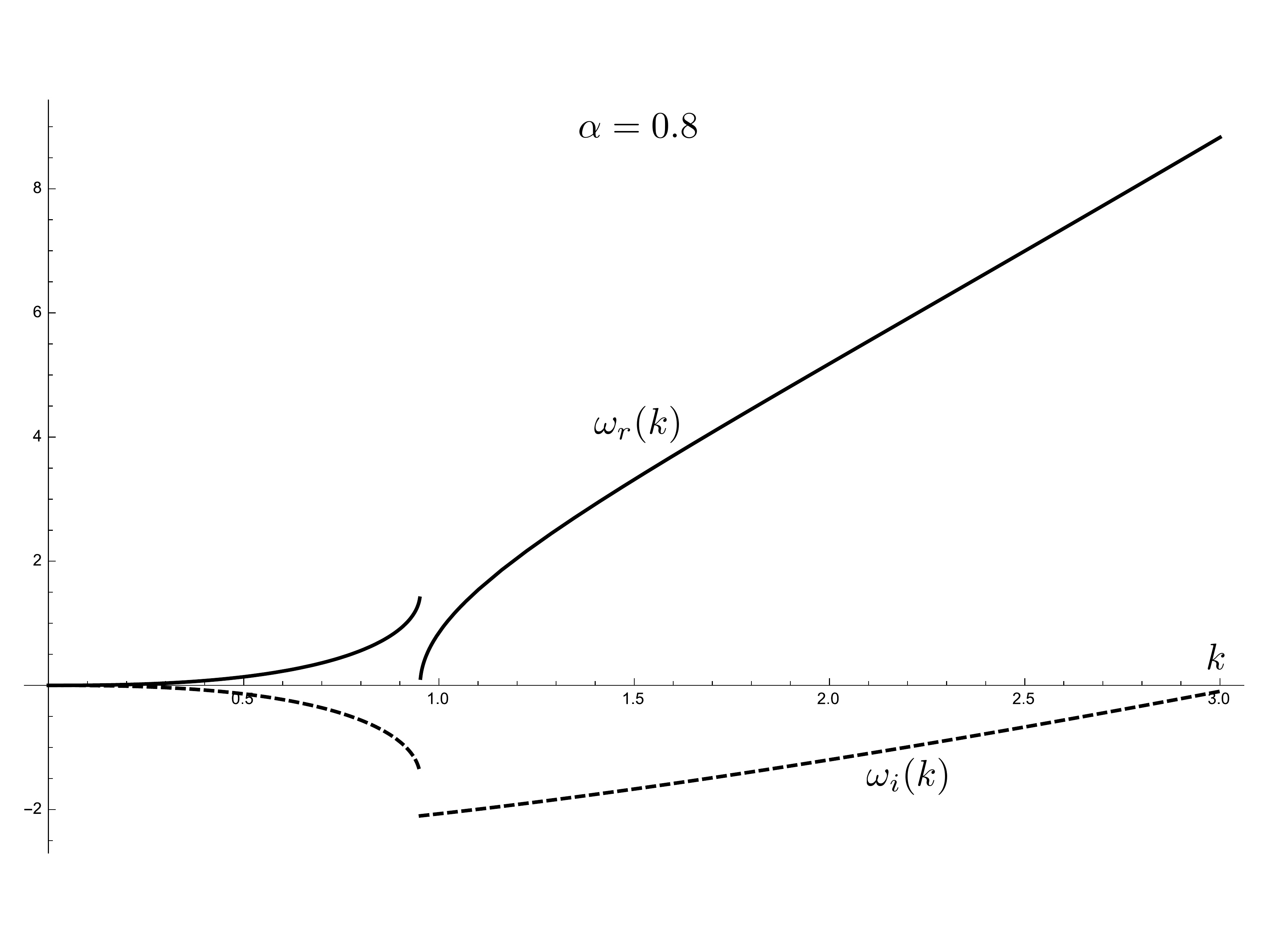}
\caption{Complex frequency as a function of the wavenumber for the fractional Cattaneo-Maxwell heat equation of order $\alpha = 4/5$. ($\tau = 0.2$ and $D = 1$) \label{fig-4}}
\end{figure}

\newpage

	From the plots one can immediately infer that, in contrast with the results in Section~\ref{sec-3}, here the time-damping $\gamma (k) = - \omega _i (k)$ is strongly $k-$dependent. Indeed, in Figure \ref{cazzafa} and \ref{fig-4} we see that for large $k$ it even turns into a forcing factor.

\vskip 0.3 cm

	We can now compute the two velocities for the time-fractional Cattaneo-Maxwell heat diffusion law. Recalling that
	$$ v_p (k) = \frac{\omega _r (k)}{k} \, , \quad 
	v_g (k) = \frac{\partial \omega _r (k)}{\partial k} \, , $$
then it is easy to see that
	\begin{equation*}
	 v_p (k) =
	\left\{	
	\begin{aligned}
	& - \frac{1}{\tau \, k} \, \left( \frac{1 \mp \sqrt{1 - 4 \, \tau ^\alpha \, \overline{D} \, k^2}}{2} \right) ^{1/\alpha} \, 
	\sin \left( \frac{\pi}{\alpha} \right) \, , \quad 0 \leq k \leq 1 / \sqrt{4 \, \overline{D} \, \tau ^\alpha} \, , \\
	& \pm \frac{1}{k} \, \left( \frac{\overline{D} \, k^2}{\tau ^\alpha} \right)^{1/2 \alpha}
	\, \sin \left[ \frac{1}{\alpha} \,\arctan \left( \sqrt{4 \, \tau ^\alpha \, \overline{D} \, k^2 - 1} \right) \right] , \quad
	k > 1 / \sqrt{4 \, \overline{D} \, \tau ^\alpha} \, ,
	\end{aligned}
	\right.	
	\end{equation*}

	\begin{equation*}
	 v_g (k) =
	\left\{	
	\begin{aligned}
	& \mp \frac{2}{\tau \, \alpha} \, \frac{\overline{D} \, \tau ^\alpha k}{\sqrt{1 - 4 \, \tau ^\alpha \, \overline{D} \, k^2}} 
	\left( \frac{1 \mp \sqrt{1 - 4 \, \tau ^\alpha \, \overline{D} \, k^2}}{2} \right) ^{\frac{1- \alpha}{\alpha}} \, \sin \left( \frac{\pi}{\alpha} \right) 
	\, ,\\
	&0 < k < 1 / \sqrt{4 \, \overline{D} \, \tau ^\alpha} \, , \\
	& \pm \frac{1}{\alpha \, k} \left( \frac{\overline{D} \, k^2}{\tau ^\alpha} \right)^{1/2 \alpha} 
	\Bigg\{ \sin \left[ \frac{1}{\alpha} \,\arctan \left( \sqrt{4 \, \tau ^\alpha \, \overline{D} \, k^2 - 1} \right) \right] \\
	& + \frac{\cos \left[ \frac{1}{\alpha}
	\arctan \left( \sqrt{4 \, \tau ^\alpha \, \overline{D} \, k^2 - 1} \right) \right]}{\sqrt{4 \, \tau ^\alpha \, \overline{D} \, k^2 - 1}} \Bigg\} \, , 
	\quad k > 1 / \sqrt{4 \, \overline{D} \, \tau ^\alpha} \, ,
	\end{aligned}
	\right.	
	\end{equation*}

\begin{figure}[h!] 
\centering
\includegraphics[scale=0.31]{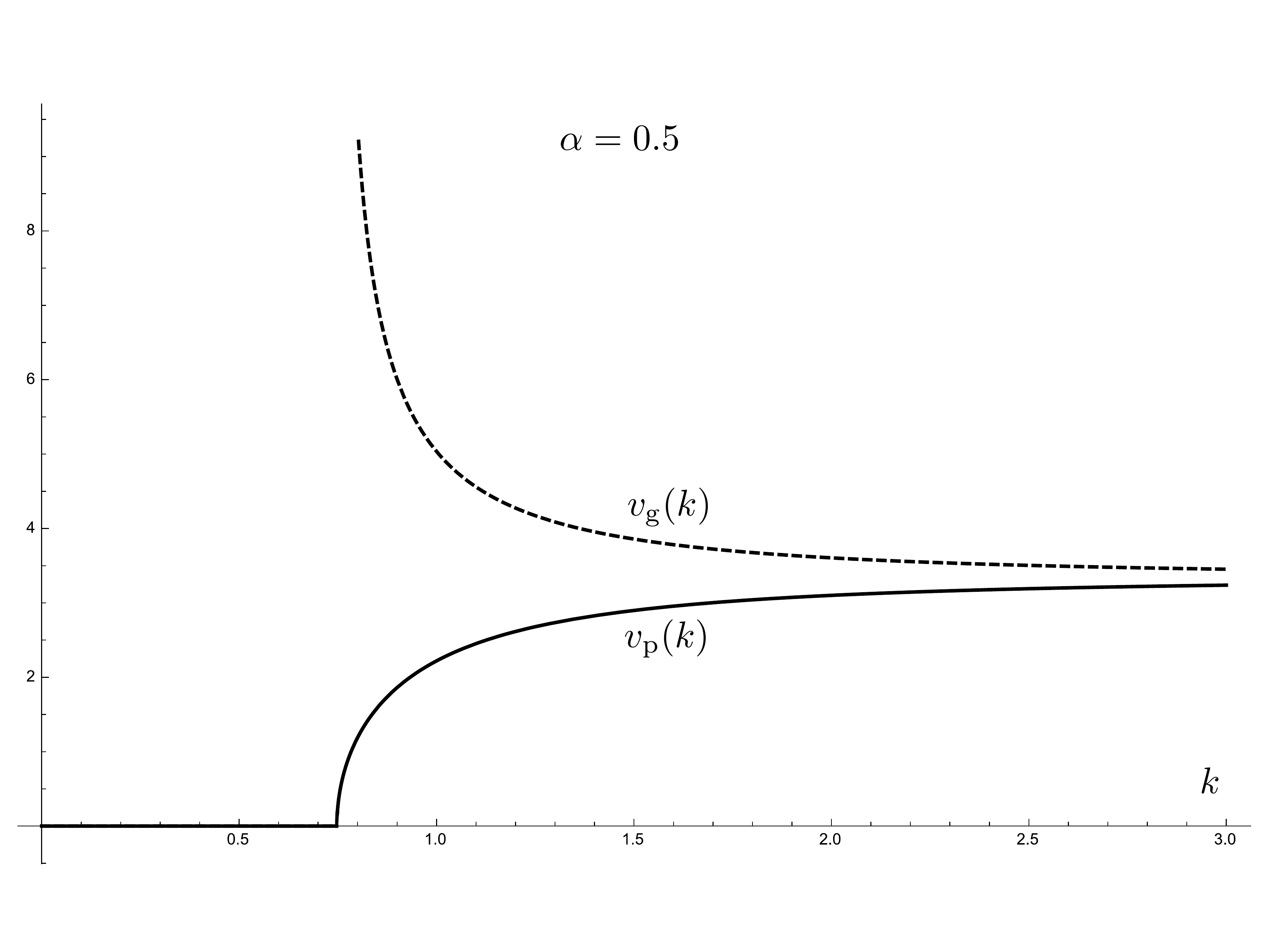}
\caption{Phase and group velocities as functions of the wavenumber for the fractional Cattaneo-Maxwell heat equation of order $\alpha = 1/2$. ($\tau = 0.2$ and $D = 1$) \label{fig-5}}
\end{figure}

\newpage

\begin{figure}[h!] \label{fig-6}
\centering
\includegraphics[scale=0.31]{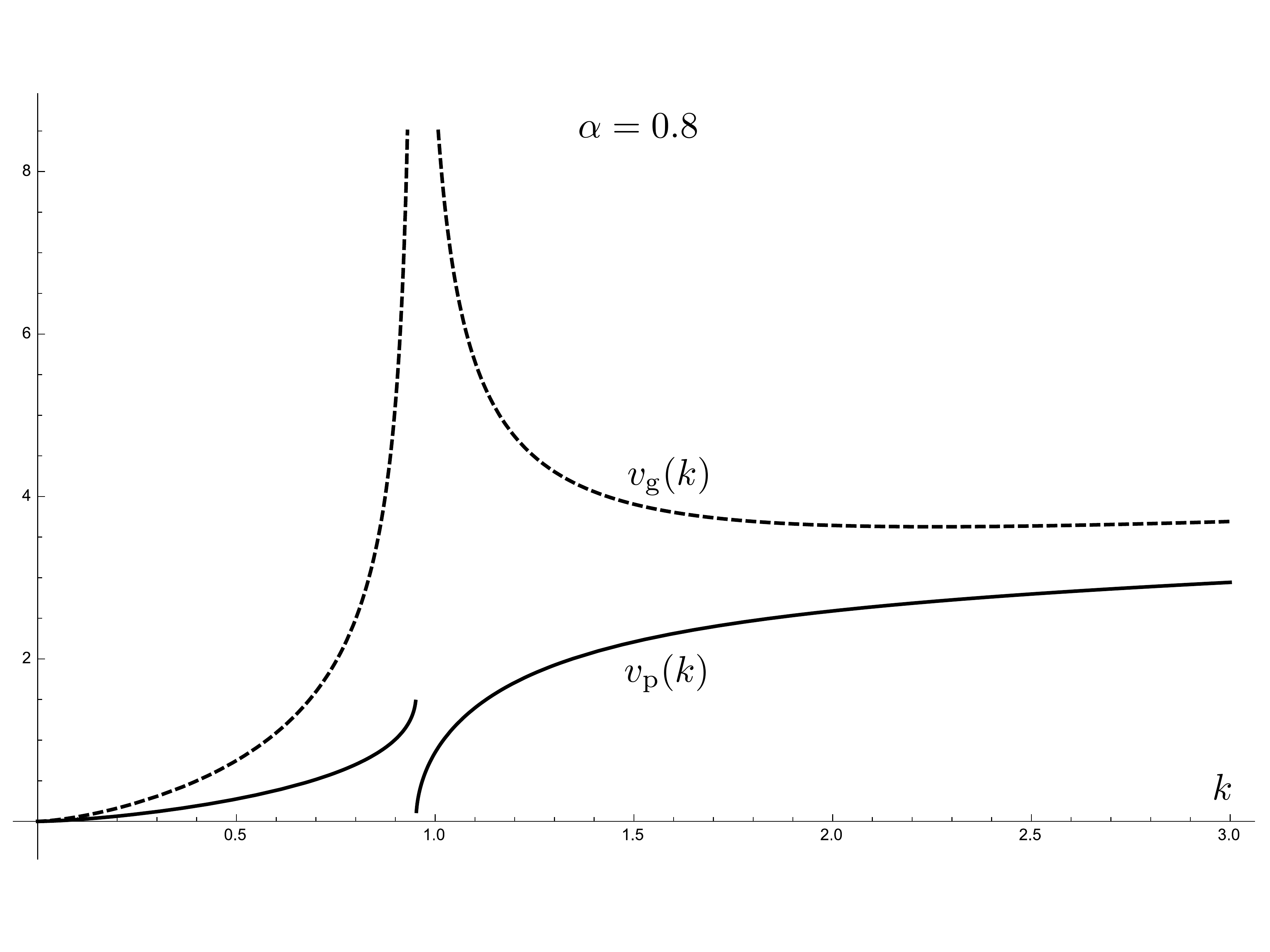}
\caption{Phase and group velocities as functions of the wavenumber for the fractional Cattaneo-Maxwell heat equation of order $\alpha = 4/5$. ($\tau = 0.2$ and $D = 1$)}
\end{figure}

\begin{figure}[h!] 
\centering
\includegraphics[scale=0.31]{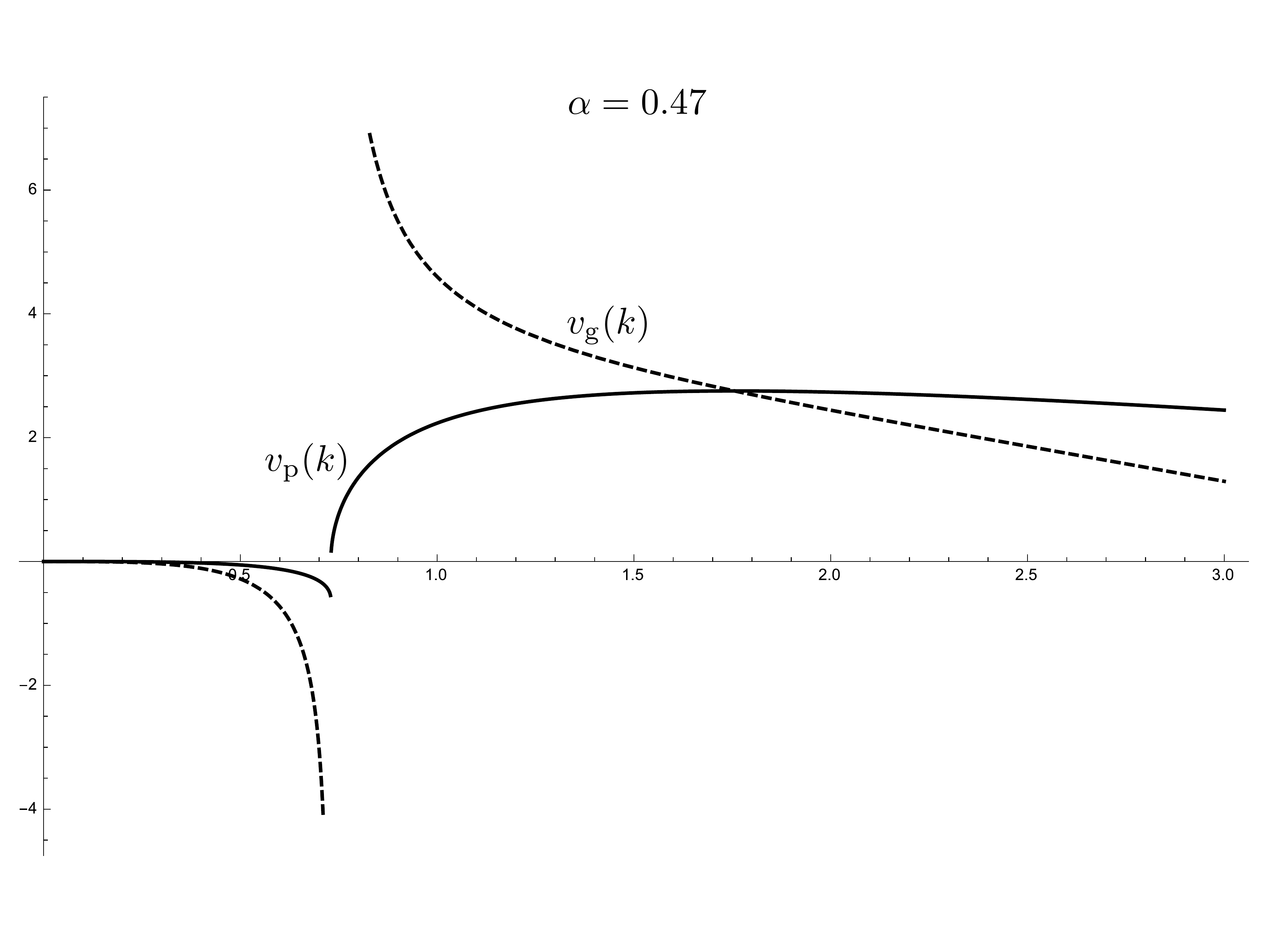}
\caption{Phase and group velocities as functions of the wavenumber for the fractional Cattaneo-Maxwell heat equation of order $\alpha = 0.47$. ($\tau = 0.2$ and $D = 1$) \label{m}}
\end{figure} 

This clearly shows that the nature of the jump discontinuity of $\omega _r (k)$ is preserved in $v_p (k)$ while an extra divergence can arise in $v_g (k)$ as we approach $a = 1 / \sqrt{4 \, \overline{D} \, \tau ^\alpha}$ from both sides. Furthermore, according to Figure~\ref{m} there are cases in which the nature of the dispersion strongly depends on $k$, indeed we have a transition from anomalous ($v_g > v_p$) to normal dispersion ($v_g < v_p$).

\section{Conclusions} 
	After a short review of some general aspects of dispersive waves with dissipation, together with a summary of the known results concerning the dispersion of waves for the causal heat propagation, a full discussion of the dispersion relation for the $(1+1)$-dimensional fractional Cattaneo-Maxwell heat conduction law is presented.
	
	Specifically, in Section~\ref{sec-4} it is observed that the fractional nature of Eq.~\eqref{FCM} can lead to discontinuities in the solutions of the dispersion law. In more details, it is argued that, unless the fractional parameter $\alpha = 1/n$ with $n \in \N$ and $n >1$, jump discontinuities would appear in both the real and imaginary parts of the complex frequency $\overline{\omega} (k)$. This result is already in strong contrast with the ordinary behaviour, for which $\overline{\omega} (k)$ is continuous and presents a cusp in $k = 1 / \sqrt{4 \, D \, \tau}$. Furthermore, in the fractional case the time-damping factor can change sign, leading to the rise of a forcing factor, depending on the value of the wavenumber.

	It is also worth noticing that some transitions from anomalous to normal dispersion may occur depending on the value of $0 < \alpha < 1$ and $k > 0$, despite what happens in the ordinary case for which we have a purely anomalous dispersion.

\section*{Acknowledgments}
	The author is thankful to Ivano Colombaro and Tommaso Ruggeri for many helpful discussions. Furthermore, the author is also deeply grateful to the anonymous referee for all the constructive comments and suggestions which have helped to significantly improve the manuscript.
	
	The work of the author has been carried out in the framework of the activities of the National Group of Mathematical Physics (GNFM, INdAM). Moreover, this work has been partially supported by \textit{GNFM/INdAM Young Researchers Project} 2017 ``Analysis of Complex Biological Systems''.


\end{document}